# Vacancy defect positron lifetimes in strontium titanate

R.A. Mackie,[1] S. Singh,[1] J. Laverock,[2] S.B. Dugdale,[2] and D.J. Keeble[1,*]

[1]*Carnegie Laboratory of Physics, School of Engineering, Physics, and Mathematics, University of Dundee, Dundee DD1 4HN, UK*

[2] *H.H. Wills Physics Laboratory, University of Bristol, Tyndall Avenue, Bristol BS8 1TL, UK*

[*]Corresponding author. Electronic address: d.j.keeble@dundee.ac.uk

The results of positron annihilation lifetime spectroscopy measurements on undoped, electron irradiated, and Nb doped $SrTiO_3$ single crystals are reported. Perfect lattice and vacancy defect positron lifetimes were calculated using two different first-principles schemes. The Sr vacancy defect related positron lifetime was obtained from measurements on Nb doped, electron irradiated, and vacuum annealed samples. Undoped crystals showed a defect lifetime component dominated by trapping to Ti vacancy related defects.

# I. INTRODUCTION

Vacancies are normally assumed to be the dominant point defects in perovskite oxide (ABO$_3$) titanates, and often strongly influence the material properties such as electrical conductivity and domain stability and dynamics.[1,2] Strontium titanate is one of the most widely studied oxide materials, having a model cubic perovskite oxide structure at room temperature and exhibits an antiferrodistortive phase transition at 105 K. The material is a wide band gap semiconductor (with $E_g^0 = 3.3$ eV,[3]) but can be made a good conductor by doping, for example, with Nb.[4] Single crystal SrTiO$_3$ substrates are readily available, and high quality epitaxial layers can be grown.[5-7] The recent observation of two-dimensional electron gas formation at the interface between SrTiO$_3$ and LaAlO$_3$,[8] residing in the SrTiO$_3$, has further demonstrated the importance of the material for future multifunctional oxide electronic device structures.[9] The electronic quality of SrTiO$_3$ is ultimately limited by the presence of electrically active point defects.

The presence and importance of vacancy-type defects in SrTiO$_3$ has been inferred from electrical conductivity measurements as a function of thermal treatments, and the Sr/Ti ratio, analyzed using defect chemistry models.[10-14] Evidence has also been presented that vacancy defects can also control the grain boundary structure, and hence grain growth.[15] The SrTiO$_3$ structure is shown in Fig. 1; the Ti$^{4+}$ ion is at the perovskite B-site and has octahedral coordination, and the Sr$^{2+}$ is the A-site ion with twelve O nearest neighbors. A good theoretical understanding of point-defect formation energies in SrTiO$_3$ has been developed using static lattice calculations with empirical and semiempirical interatomic potentials,[16, 17] and first principles calculations.[18] These studies show that Schottky disorder ($V_{Sr}^{2-} + V_{Ti}^{4-} + 3V_O^{2+}$) has a lower formation energy than Frenkel-type reactions involving interstitials. It was also found that the Sr partial Schottky defect ($V_{Sr}^{2-} + V_O^{2+}$) could have an even smaller formation energy.[16] A detailed first principles study as a function of Fermi level, at a number of points in the phase diagram, has confirmed the potential importance of the Sr partial Schottky defect compared to the full, and Ti partial, Schottky defects, and the authors have also calculated formation energies for the isolated vacancy defects as a function of the charge state varying from the ionic value to the neutral state.[18] In addition, Tanaka *et al.*[18] report the atomic and electronic structures of the isolated vacancy defects and have provided detailed information on structural relaxations.

Evidence for the presence of vacancy-type defects in SrTiO$_3$ has also been provided by tracer diffusion measurements using appropriate isotopes of Sr, Ti, or O.[19-21] Direct imaging of O vacancies has been demonstrated, when present at sufficiently high concentrations, using aberration corrected transmission electron microscopy (TEM),[22] or annular-dark-field TEM, supported by electron energy-loss spectroscopy (EELS).[23] Electron energy-loss near-edge structures (ELNES) combined with first-principles calculations provided evidence for high concentrations of Sr vacancies in the vicinity of grain boundaries, resulting from heat treatments.[24] However, there has not been an unambiguous direct observation of vacancy defects in a perovskite oxide titanate using atomic scale spectroscopy methods, such as electron magnetic resonance, that can provide information on local structure. Positron annihilation spectroscopy (PAS) techniques are capable of providing atomic scale structural information on vacancy-related point defects in materials. The positron is strongly localized at open-volume defects, where the lifetime of the positron increases.

Here positron annihilation lifetime spectroscopy (PALS) experiments on a series of SrTiO$_3$ single crystals from several different suppliers are presented. These included as-received, Nb-doped, vacuum annealed, and electron irradiated samples. In addition, the results of positron lifetime calculations for perfect SrTiO$_3$, and various vacancy-type defect configurations, using first-principles density functional theory calculations are given.

Positrons are readily obtained from suitable radioisotopes, for example $^{22}$Na which is used in this study, and are implanted to a distribution of depths determined by the emission energy spectrum of the isotope, the material density and atomic number. Typically the maximum implantation depth is of order 0.1 mm. The implanted positron thermalizes within a few picoseconds. It then annihilates in the material from a state $i$ with a lifetime $\tau_i$ and a probability $I_i$. This can be a delocalized state in the bulk lattice, or a localized state at a vacancy defect, or can in certain circumstances (*e.g.* in nanovoids) form the electron-positron bound state known as positronium. If the average lifetime, $\bar{\tau} = \sum_i I_i \tau_i$, is greater than the bulk lattice lifetime characteristic of the material, $\tau_B$, then it indicates that vacancy-type defects are present. The rate of positron trapping to a vacancy, $\kappa_d$, is proportional to the concentration of these defects, $[d]$, where the constant of proportionality is the defect specific trapping coefficient, $\mu_d$. The one defect trapping model (1D-STM) gives,

$$\kappa_d = \mu_d [d] = I_2 \left( \frac{1}{\tau_1} - \frac{1}{\tau_2} \right), \quad (1)$$

predicting two experimental lifetimes; $\tau_2 = \tau_d$ is the characteristic defect lifetime, and the first lifetime, $\tau_1$, is reduced from the bulk lifetime, $\tau_B$, by an amount that depends on the rate of trapping to the defect, such that $\tau_1 < \tau_B$. If the 1D-STM is applicable the bulk lifetime can be calculated from the experimental lifetime components,

$$\tau_B = \left( \frac{I_1}{\tau_1} + \frac{I_2}{\tau_2} \right)^{-1} \quad (2).$$

The model is readily extended to include more defect states, and can also allow for detrapping from shallow states.[25] As the concentration of defects increases, the intensity of the longer lifetime, $I_2$, increases towards unity. Eventually, all positrons will annihilate from the defect. This is known as saturation trapping and sensitivity to defect concentration is restricted or lost completely; the concentration at which this occurs depends on the specific values of $\tau_B$ and $\mu_d$. Assuming a plausible specific trapping coefficient ~ $2 \times 10^{15}$ s$^{-1}$ at. for a negatively charged monovacancy,[25] saturation can be estimated to occur when $\kappa_d \tau_B = \mu_d [d] \tau_B \approx 10$, resulting in a vacancy concentration of the order of 50 ppm. Positron lifetime spectroscopy has the advantage over other PAS methods in allowing different positron states to be simultaneously resolved. However, the width and accuracy of the experimental instrument function imposes a limit; typically a lifetime component must be at least ~40 % longer than the previous one to allow them to be separated, or else a weighted average of the unresolved components is detected.

Early positron lifetime measurements on Bi-doped,[26, 27] and La-doped,[27-29] SrTiO$_3$ reported a defect lifetime in the range ~ 360–510 ps, the high values of the first lifetime and the large simple trapping model-calculated bulk lifetimes show that saturation trapping to defects had occurred. The results were discussed in terms of isolated $V_{Sr}^{2-}$, defect complexes including $La_{Sr}^+ V_{Sr}^{2-}$ and $2La_{Sr}^+ V_{Sr}^{2-}$, and trapping at grain boundaries. However, their interpretation was hampered by the lack of available theoretical positron lifetimes. A nanosecond orthopositronium pick-off lifetime was also detected from La doped ceramics, providing evidence for the presence of nanovoids.[27, 28]

More recently the atomic superposition method has been used to calculate the lifetimes of both delocalized and trapped positron states in perovskite oxide materials.[30-32] These provide evidence that the bulk positron lifetime for annihilation from a perfect lattice is in the range ~ 130–160 ps, the isolated B-site vacancy is ~ 170–205 ps, the more open A-site vacancy range is ~ 240–295 ps, and the O vacancy lifetime value is slightly longer than the bulk at ~ 140–165 ps. Formation of O vacancy – cation vacancy complexes increases the lifetime with respect to the single cation vacancy. These results suggest that the lifetime values in the range 360–510 ps observed in the previous studies of ceramic SrTiO$_3$ resulted from open volume trapping sites larger than A-site vacancy defects.

Positron lifetime measurements in SrTiO$_3$ single crystal samples have also been reported.[31, 33, 34] A single lifetime was found for Nb-doped crystals of 141(1) ps at room temperature,[34] and varying in the range ~130–140 ps between 20 to 350 K.[33] A measurement on similar samples used to grow 30 nm BaTiO$_3$ thin films by molecular beam epitaxy (MBE) then exhibited a defect lifetime at ~ 294 ps.

## II. POSITRON LIFETIME CALCULATIONS

Calculations of the positron lifetime were performed using two different density functional theory (DFT) methods: one using the MIKA/Doppler package,[35] and the other with the linear muffin-tin orbital (LMTO) method.[36] In the MIKA/Doppler package, the electron density of the solid is approximated by the non-self-consistent superposition of free atom electron densities in the absence of the positron (the so-called 'conventional scheme').[37] This approximation to the complete two-component density functional theory (TCDFT) has been found to give positron lifetimes close to TCDFT as well as experimental values. In our calculations of the positron lifetimes, two schemes for the electron-positron enhancement factor (that accounts for the enhancement of the electron density at the positron) have been used, that due to Barbiellini and co-workers,[38, 39] as a parameterization of the data of Arponen and Pajanne,[40] (referred to as AP) and that due to Boronski and Nieminen (BN),[41] both of which have been described within the generalised gradient approximation (GGA) of the enhancement factor. Calculations were performed with the room temperature Pm3m structure of SrTiO$_3$,[42, 43] using a 6×6×6 supercell.

Calculations that included the relaxation of neighbor atoms surrounding the isolated vacancies were also made.

Additionally, self-consistent calculations of the electron density (where the charge density is iteratively computed until self-consistency is achieved) have been performed using the LMTO method within the atomic sphere approximation, and including combined correction terms.[36] These calculations used a basis of *s*, *p* and *d* states for the wavefunctions of the occupied sites, and a reduced basis of *s* and *p* states for the vacancies. Calculations were performed using both 2×2×2, as well as 3×3×3, supercells, and self-consistence in the charge density was achieved at 27 k-points (for the 2×2×2 supercell) and 8 k-points (for the 3×3×3 supercell) in the irreducible Brillouin zone (1/48th of the full Brillouin zone). The positron lifetime was then computed using both the AP and BN enhancement factors, again employing the GGA.

## III. EXPERIMENT

Undoped $SrTiO_3$ single crystals were obtained from a number of different suppliers, including MaTecK GmbH, Shinkosha Co. Ltd, MTI Corporation, Escete BV, and PI KEM Ltd. Niobium-doped $SrTiO_3$ crystals from two suppliers were used. Electron irradiation was performed using the linear accelerator (LINAC) at the National Physical Laboratory. Two sets of MaTecK $SrTiO_3$ substrates were subjected a ~3 MeV electron beam and received doses of $1\times10^{17}$ $e^-/cm^2$ and $3\times10^{18}$ $e^-/cm^2$, respectively; the sample temperature was monitored and did not exceed 60° C. Vacuum annealing experiments were performed using a movable tube furnace connected with a base pressure better than $2\times10^{-6}$ Pa.

The positron lifetime experiments were performed using two conventional fast-fast spectrometers. One is optimized for room temperature measurements and with a time resolution function of 205 ps, the second for variable temperature studies in the range 12-300 K with a resolution function of 235 ps.[44] A detailed description of the resolution function required for spectrum analysis used three Gaussian functions with intensities 80%, 10%, 10%, and appropriate relative shifts. All spectra contained at least $5 \times 10^6$ counts. Positron sources were made from aqueous $^{22}$NaCl either directly deposited on the material under study or deposited on thin support foils, either 1$\mu$m Ni, 8$\mu$m Kapton or 1.5$\mu$m Al, with source strengths between 200 to 850 kBq. Two near identical samples sandwiched the positron source. The lifetime spectrum is analyzed as a sum of exponential decay components, $n(t) = \sum_i I_i \exp(-t/\tau_i)$, convoluted with the Gaussians functions describing the spectrometer timing resolution using POSITRONFIT.[45] Decay components due to annihilations in the positron source were subtracted in the procedure, direct deposit sources required one component with a lifetime of ~430 ps with an intensity that depended on source strength, foil supported sources required an additional component with a lifetime of ~380 ps for Kapton (~10 %), 120 ps for Ni (~10 %), or 190 ps for Al (~5 %). Care is required to correctly account for source annihilations, measurements were made with at least two types of foil source to establish source independent fittings.[44] This procedure involved a systematic variation of the source correction terms and a minimization of the fit chi-squared and allowed both material and source correction terms to be determined and checked.

Studies were also performed using simulated spectra. These were generated from the relevant simple trapping model using the calculated positron lifetimes, and assumed values for positron trapping coefficients and the defect concentrations. The resulting sum of exponential decays was then convolved with the experimental spectrometer timing resolution function, noise added and the area normalized to simulate a $5 \times 10^6$ count spectrum. These spectra were then analyzed using POSITRONFIT.

## IV. RESULTS AND DISCUSSION

The calculated positron lifetime values for the bulk and for the relevant vacancy defects in $SrTiO_3$ are given in Table I, along with previously reported values.[31] The LMTO calculation results shown were made using 3×3×3 supercells.

Comparing the results from the present atomic superposition method using the Boronski and Nieminen (BN) positron enhancement factor with the previously reported values,[31] given in Table I, shows there is approximate agreement for the unrelaxed structures. The agreement is good for Sr vacancy defects, and is reasonable for the perfect lattice (bulk) values. The LMTO BN bulk value is significantly smaller. Calculations made using the AP positron enhancement give longer lifetime values, and have been found to provide good agreement with experiment for the Pb-based perovskite oxides.[32] The different calculations show the same trend in the positron lifetime as the positron state is varied from delocalized bulk to trapped states at anion or cation vacancies. The calculated bulk lifetime values span the range 121–151 ps, the $V_O$ 133–166 ps, the B-site vacancy 156–195 ps, and A-site vacancy 238–279 ps (Table I). Similar trends, and approximate values, have been reported for other perovskite oxides.[30, 32] The table also includes some calculations using relaxed defect structures, obtained from the *ab-initio* calculations. The relaxed structures for the Ti vacancy and the Sr vacancy were taken from Tanaka *et al.*,[18] and those for the oxygen vacancy from Carrasco *et al.*,[46] which can be compared to the results of Hamid *et al.*.[31] The inward relaxation for $V_{Ti}$ causes a reduction in the lifetime, but there is essentially no effect for the more open $V_{Sr}$.

The formation of complexes between cation vacancies and nearest neighbor charge compensation oxygen vacancies results in a marked increase in the calculated lifetime at the 6-coodinated B-site, typically between 20–30 ps, but the effect is much smaller at the more open 12-coordinated A-site. However, the Sr site may allow the formation of larger $V_A$-$n V_O$ complexes (Fig. 1),[47] which then result in a noticeable increase in lifetime compared to the isolated vacancy (Table I).

The experimental room temperature positron lifetimes for a series of commercially available, nominally undoped, single crystal $SrTiO_3$ samples are shown in Fig. 2, and those for Nb-doped, electron irradiated, and vacuum annealed, single crystal $SrTiO_3$ and ceramic $SrTiO_3$ are given in Fig. 3. Differences were observed between different batches from the same supplier (Fig. 2). Apart from Nb-doped samples from one supplier, all analyzed lifetime spectra showed significantly smaller variances (< 1.10) for two lifetimes-component fits, and relevant results are summarized in Table II.

The positron lifetime for annihilations from perfect SrTiO$_3$, $\tau_B$, is of importance for the quantitative analysis of positron annihilation spectra. In principle if the correct number of positron states are included in the simple trapping model (STM), the calculated bulk lifetime (*e.g.* Eq. (2)) should be a constant for all SrTiO$_3$ samples studied and should be equal to the true bulk value. The bulk value can not be obtained from samples exhibiting saturation trapping to defect states. If the sample is not in saturation trapping, *i.e.* $\tau_1 \leq \tau_B$, the STM calculated value may be in error if a component intensity or lifetime is incorrectly determined, for example due to errors in the source correction. It has been found that typically these errors result in $\tau_{B(STM)} \geq \tau_B$, however, if a trapped state population of small intensity is not resolved $\tau_{B(STM)} < \tau_B$ results. Correct determination of experimental components, even if resulting from the weighted average of two or more unresolved material lifetimes, should allow $\tau_{B(STM)} = \tau_B$ to be obtained.

Inspection of the experimental $\tau_{B(STM)}$ given in Table II shows the lowest value, 149 ps, was obtained from Nb-doped SrTiO$_3$ single crystals, derived from a first lifetime of 131(1) ps and vacancy defect lifetime of 272(4) ps. Niobium-doped SrTiO$_3$ single crystals from a second source were also studied and found to give a similar low $\tau_1$ value and vacancy defect lifetime, but showed a third, low intensity, ~ 1 ns orthopositronium pick-off component, most likely due to the presence of nanovoids. The average bulk value obtained from the undoped crystals was slightly longer, 155(4) ps. Previous positron lifetime measurements on Nb-doped SrTiO$_3$ single crystals have reported single component fits, with a lifetime of 141(1) ps at room temperature,[34] and 134(2) at 30 K. [31] These results suggest the experimental room temperature bulk lifetime lies in the range 140–155 ps, consistent with the values calculated using the AP positron enhancement factor (Table I).

The results from the undoped SrTiO$_3$ single crystals are given in Fig. 2 and Table II. A number of samples showed a second lifetime component, $\tau_2$, of ~202 ps [Fig. 2(a)], slightly longer than AP enhancement calculation values for the Ti vacancy, 184–195 ps, and shorter than the V$_{Ti}$ – oxygen vacancy complex values of 225–239 ps (Table I). Three possible models are considered; (i) the component is due to V$_{Ti}$ only, and theoretical calculations underestimate the lifetime, (ii) it is due to an unresolved weighted average of V$_{Ti}$ and V$_{Ti-O}$ defect populations, or (iii) it results from V$_{Ti}$ and a low concentration of V$_{Sr}$ defects. Assuming the V$_{Ti}$ lifetime is ~190 ps, the limit for deconvoluting a second lifetime is ~ 266 ps, putting the V$_{Sr}$ value close to the limit of resolution. To further investigate the experimental limits on the deconvolution, a study with simulated spectra was performed with the calculated lifetimes (MIKA-AP, Table I) and the two defect simple trapping model. An example of the results obtained using model (iii) is shown in Fig. 4. The defect-specific trapping coefficients for V$_{Ti}$ and V$_{Sr}$ were taken to be $2 \times 10^{15}$ s$^{-1}$ at. and $1 \times 10^{15}$ s$^{-1}$ at., respectively, and the concentration of Ti vacancies fixed at 0.9 ppm. The [V$_{Sr}$] was then varied, and the lifetime spectra predicted by the simple trapping model with the fitting results are plotted. They show that the two defect lifetimes were not resolved; the two term fits had best variances, with the second component containing contributions from both $V_{Ti}$ and $V_{Sr}$. The effect of increasing [$V_{Sr}$] is to increase the lifetime and intensity of this second component. With suitable choices of concentrations and trapping rates, model (ii) can also reproduce the experimental spectra. A smaller set of as-received samples showed a longer second lifetime component, see Fig. 2(b), these results can also be fitted to model (iii) with appropriate defect concentration values. A necessary condition for a defect population model consistent with the commonly observed ~202 ps lifetime component, within the simple trapping model, is the presence of a positron trap with lifetime equal to, or less than, this value. This restricts the possible candidates to the O or Ti monovacancies, or the oxygen divacancy defect (Table I).

Variable temperature positron lifetime measurements were performed on a typical undoped (MaTecK GmbH) $SrTiO_3$ sample and the results are shown in Fig. 5. The initial second component lifetime of 199(4) ps reduced systematically with reducing temperature toward a low temperature value of ~ 181 ps, the associated intensity increased. At 40 K and lower only a single lifetime 181(3) ps fit was required. These results are consistent with an increasing trapping coefficient with decreasing temperature, as expected for a negatively charged vacancy defect and opposite to that for neutral vacancies.[48] The observations provide evidence that positron trapping is increasingly dominated by a negatively charged vacancy defect with a lifetime of 181(3) ps at low temperatures. The Ti vacancy is the only negative vacancy with a sufficiently short lifetime and the agreement with the relaxed structure MIKA-AP calculation value of 184 ps is good (Table II).

To further investigate this model simulations were performed. The best fit to the 293 K spectrum required a two-defect trapping model simulation, and the resulting POSITRONFIT contained two terms and is shown in Fig. 5(c) ([$V_{Sr}$] = 0.35 ppm, $\tau(V_{Sr}) = 280$ ps, $\mu(V_{Sr}) = 1 \times 10^{15}$ s$^{-1}$at., [$V_{Sr}$] = 0.90 ppm, $\tau(V_{Ti}) = 184$ ps, $\mu(V_{Ti}) = 2 \times 10^{15}$ s$^{-1}$at.). The trapping coefficient assigned to $V_{Ti}$ was then systematically increased, comparison with the experiment (Fig. 5(b)) shows the trend with decreasing temperature is in approximate agreement. This provides further evidence for the assignment of the $V_{Ti}$ lifetime to ~184 ps, and is consistent with the value for the B-site vacancy inferred from experiments on Pb(Zr$_{0.42}$Ti$_{0.58}$)O$_3$ ceramics.[32]

Figure 3 shows the experimental lifetime values obtained from Nb-doped, electron-irradiated, and vacuum-annealed, single crystal SrTiO$_3$ and ceramic SrTiO$_3$. A characteristic of these results was the observation of a $\tau_2$ in the range 260–280 ps. The values lie in the range expected for positron trapping to Sr-vacancy-related defects, namely 238–289 ps (Table II). The Nb-doped SrTiO$_3$ spectrum, showed a reduced bulk first lifetime and a second defect lifetime 272(4) ps due to A-site vacancies, and is markedly different from the undoped single crystal (Table II). The result provides evidence that Nb$^{5+}$ doping suppresses the Ti vacancy defect content observed from the undoped crystals, and that Sr vacancy defects contribute to charge compensation. An approximate Sr vacancy concentration was calculated assuming trapping coefficients in the range $0.5 \times 10^{15}\,s^{-1}$ at. to $6 \times 10^{15}\,s^{-1}$ at., corresponding to neutral and double negatively charge monovacancies in a semiconductors,[25] and values in the range ~0.2–2 ppm were obtained. A comparable lifetime value, 276(6) ps, was obtained from high quality undoped ceramic samples where saturation trapping was observed with the dominant first lifetime component 176(2) ps. This is consistent with contributions from $V_O$ and $V_{Ti}$ related defects.

The second resolved lifetime for the $3 \times 10^{18}$ e$^-$/cm$^2$ irradiated sample was 261(5) ps, and is comparable to that observed for the vacuum-annealed samples, 262(6) ps (Fig. 3). These are slightly shorter than the above, and may indicate that the component results from more than one vacancy defect type. The as-received MaTecK GmbH substrates used for electron irradiation showed lifetime spectra similar to that given in Table II. Displacement thresholds are not available for SrTiO$_3$, however, vacancy formation on each sublattice, typically O and Sr vacancies have lower formation energies compared to Ti vacancies.[18] The observed spectrum from the electron-irradiated sample could be explained by increased [$V_{Sr}$], compared to [$V_{Ti}$], see Fig. 4. The second lifetime obtained from the $1 \times 10^{17}$ e$^-$/cm$^2$ irradiated sample was 230 ps ($I_2 = 40\ \%$), supporting an increase in $V_{Sr}$ defects with respect to $V_{Ti}$ with dose. The same type of as-received crystals were also used for the vacuum anneal experiments, where an increase in oxygen vacancy related defects is expected. It is proposed that the change in the population distribution between vacancy types, combined with the effects of limited instrument resolution, result in the detection of a second lifetime of 262(6) ps due to unresolved components containing a significant contribution from Sr vacancy trapping.

Both electron irradiation and vacuum annealing were observed to have a marked effect on the first lifetime component (see Table II and Fig. 3). It increased from a reduced bulk value of 124(3) ps to 166(1) ps ($I_1$ = 79 %) for the $3\times10^{18}$ e$^-$/cm$^2$ sample, and to 144(1) ps ($I_1$ = 81 %) for a vacuum-annealed sample. These results are consistent with the introduction of positron trapping defects with lifetimes similar to, or slightly greater than, the bulk lifetime suggesting trapping to O monovacancy or divacancy defects.

## V. CONCLUSIONS

The positron lifetimes calculated using both the MIKA/Doppler package and the LMTO method in $SrTiO_3$ are in approximate agreement, and provide evidence that the bulk lifetime, $\tau_B$, is in the range 121–151 ps. The experimental positron lifetime measurements on Nb-doped and undoped single crystal $SrTiO_3$ gave values for the bulk lifetime calculated using the simple trapping model in the range 149–155 ps. An experimental lifetime in the range 260–276 ps observed from electron irradiated, Nb-doped, ceramic, and vacuum annealed $SrTiO_3$ is in agreement with the 238–289 ps range calculated for the perovskite A-site vacancy, and its complexes with nearest neighbor oxygen vacancies. The presence of unresolved contributions from other vacancy defects may be responsible for the lower values observed from electron irradiated and vacuum annealed samples, suggesting the isolated Sr vacancy lifetime is ~ 275 ps. A vacancy defect component lifetime in the range 198 – 232 ps was observed from undoped $SrTiO_3$ crystals. Fitting of simulated spectra that included the experimental resolution function showed this component could result if both Ti and Sr vacancies were present. At low temperatures a single lifetime component at 181(3) ps was observed, consistent with trapping to negatively charged vacancy defect. These observations, and the calculated positron lifetime values for $V_{Ti}$, provide good evidence that the dominant positron trap in the undoped samples is the Ti vacancy with an associated lifetime of ~ 181 ps.

## ACKNOWLEDGEMENTS

RAM acknowledges the support of a Carnegie Scholarship and a Research Grant from the Carnegie Trust for the Universities of Scotland. SBD would like to thank the Royal Society and the EPSRC. DJK thanks M. McEwen and A R Du Sautoy, Centre for Acoustics and Ionising Radiation, National Physical Laboratory for the electron irradiations. DJK and RAM also thank S. Stemmer (University of California, Santa Barbara) and T. J. Jackson (University of Birmingham) for $SrTiO_3$ crystal samples.

**FIGURE CAPTIONS**

FIG. 1.  (color online) a) Room temperature SrTiO$_3$ structure b) A-site vacancy with one nearest neighbor oxygen vacancy.

FIG. 2.  (color online) Positron lifetime measurements for a series of as-received undoped SrTiO$_3$ crystals. Measurements using Ni (solid), Kapton (open), Al (half fill) foil supported positrons sources are shown, directly deposited positron sources (gray fill) were also used. (a) crystals with similar $\tau_2$ values (b) other as-received crystals.

FIG. 3.  (color online) Positron lifetime measurements on electron irradiated (down triangle), Nb-doped (circle) and two pairs of vacuum annealed (diamond and square) SrTiO$_3$ crystals are shown along with ceramic SrTiO$_3$ (up triangle). Measurements using Ni (solid), Kapton (open), Al (half fill) foil supported positron sources are shown.

FIG. 4.  (color online) The results of fitting simulated positron lifetime spectra (solid symbols) obtained by assuming the presence of Ti vacancies and Sr vacancies, assuming lifetime values calculated using MIKA-AP (Table I) and with trapping coefficients values of $2 \times 10^{15}$ s$^{-1}$ at. and $1 \times 10^{15}$ s$^{-1}$ at., respectively. The [V$_{Ti}$] was fixed at 0.9 ppm and the [V$_{Sr}$] varied; 0.35 (up triangle), 0.9 (circle), 4 (down triangle), and 9 ppm (square). The simple trapping model spectra used to construct the simulated spectra are also shown (open symbols).

FIG. 5. (color online) (a) Average positron lifetime as a function of temperature for undoped SrTiO$_3$ single crystal samples (MaTecK GmbH), (b) experimental two component positron lifetime fit results, (c) simulation results showing the two defect trapping model components (open symbols) used to generate the simulated spectrum, and the resulting POSITRONFIT fits (solid symbols). The trapping coefficient of V$_{Ti}$ was varied. A single lifetime fit was found for 5 x 10$^{16}$ s$^{-1}$ at. (model components, $\tau_1$ = 19.2 ps, I$_1$ = 3.3%, I$_2$ = 96.0%, I$_3$ = 0.7%).

## TABLES AND TABLE CAPTIONS

TABLE I. Calculated positron lifetime values (ps) for SrTiO$_3$. The MIKA and LMTO calculations performed here using different positron enhancement factors and with values obtained using relaxed structures from Ref. [18] and [46] given in parentheses. These are compared to atomic superposition calculations given in Ref. [31] (relaxed structure calculations in parentheses).

|  | MIKA BN | MIKA AP | LMTO BN | LMTO AP | AT-SUP BN Ref. [31] |
|---|---|---|---|---|---|
| Bulk | 131 | 151 | 119 | 146 | 138 |
| $V_O$ | 142 (146) | 166 (170) | 145 | 159 | 141 (164) |
| $V_{O-O}$ | 155 | 178 |  |  | 145 (180) |
| $V_{Ti}$ | 160 (157) | 195 (184) | 174 | 194 | 170 |
| $V_{Ti-O}$ | 188 | 225 |  |  | 199 |
| $V_{Sr}$ | 238 (238) | 279 (279) | 227 | 252 | 240 |
| $V_{Sr-O}$ | 244 | 283 |  |  | 246 |
| $V_{Sr-2O}$ | 250 | 286 |  |  |  |
| $V_{Sr-3O}$ | 256 | 289 |  |  |  |

TABLE II. Experimental positron lifetime values (ps) for SrTiO$_3$ samples, they are single crystal unless state, either undoped (UD), Nb-doped, or electron irradiated (dose $3\times10^{18}$ e$^-$ cm$^{-2}$). The average lifetimes and the one defect simple trapping model bulk values, obtained using Eq. (2), are also given.

| Sample | | $\tau_1$ | $\tau_2$ | $I_2$ (%) | $\bar{\tau}$ | $\tau_{B(STM)}$ |
|---|---|---|---|---|---|---|
| Shinkosha | UD | 130.6(2.0) | 201.3(2.9) | 41.2(3.2) | 160 | 153(7) |
| MaTecK | UD | 123.5(3.3) | 201.3(2.5) | 54.3(3.5) | 166 | 156(9) |
| MTI | UD | 120.0(2.5) | 205.8(2.0) | 55.6(2.4) | 168 | 156(6) |
| PI-Kem | UD | 94.4(5.5) | 198.5(2.5) | 77.9(3.1) | 175 | 159(10) |
| MaTecK | UD | 128.3(1.7) | 222.4(3.0) | 42.6(2.5) | 168 | 156(6) |
| Toplant | Nb | 130.6(9.0) | 272.4(4.0) | 23.4(1.0) | 164 | 149(9) |
| MaTecK | e$^-$ | 166.3(1.3) | 260.7(4.7) | 20.9(2.0) | 186 | 180(5) |
| Ceramic | UD | 176.3(2.0) | 276.3(6.0) | 6.2(1.8) | 183 | 180(4) |

FIG. 1

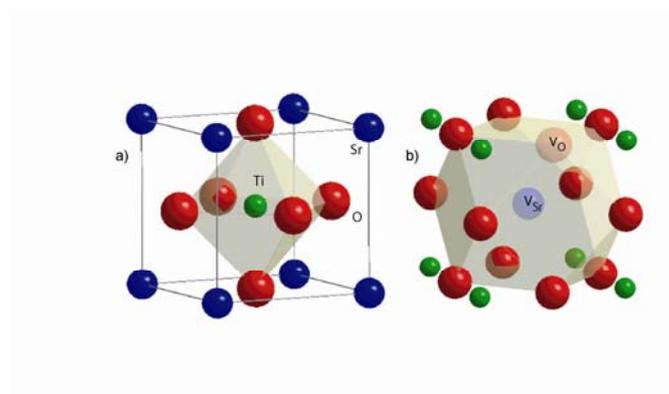

FIG. 2

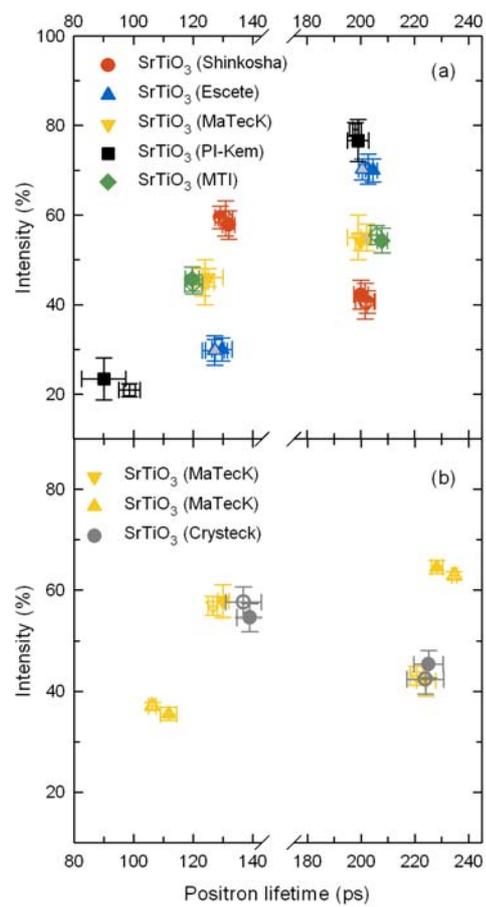

FIG. 3

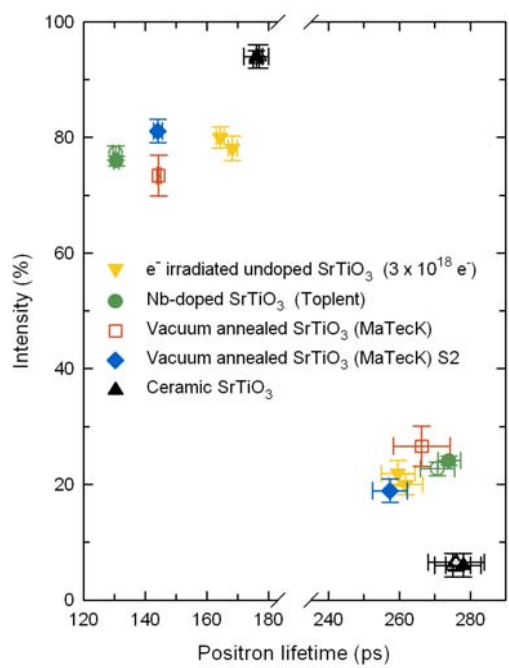

FIG. 4

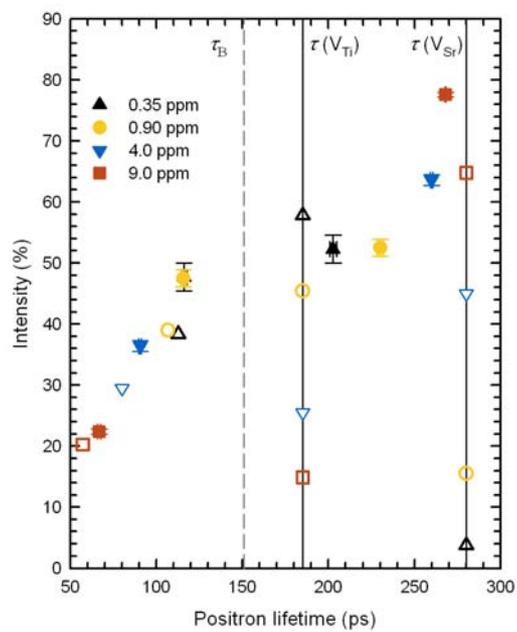

FIG. 5

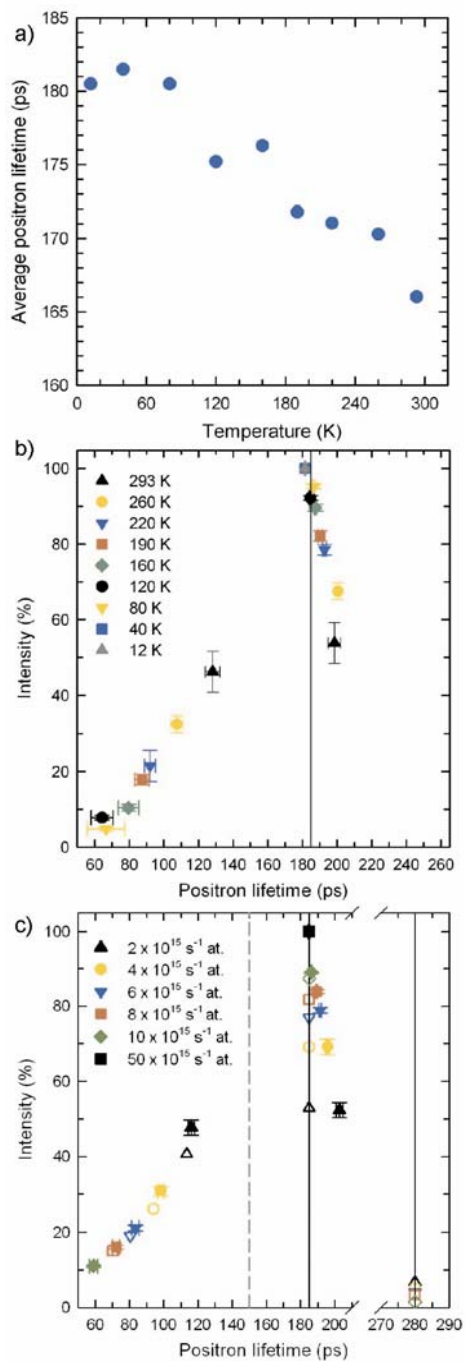